\documentclass[]{JFM-FLM_Au}

\usepackage{graphicx}
\graphicspath{{figures/}} 
\usepackage{dcolumn}
\usepackage{bm,comment}
\usepackage[utf8]{inputenc}
\usepackage[T1]{fontenc}  
\usepackage[dvipsnames]{xcolor}

\renewcommand{\Pr}{Pr}
\newcommand{\Sc}{Sc}
\newcommand{\Ra}{\mathcal{R}}
\newcommand{\R}{\varepsilon}
\newcommand{\fslow}[1]{\left\langle #1 \right\rangle_f}
\newcommand{\ffast}[1]{\tilde{#1}}

\lefttitle{A.E.~Fraser, \textit{et al.}}
\righttitle{Journal of Fluid Mechanics}

\title{Spontaneous generation of helical flows by salt fingers}

\author{Adrian E. Fraser\aff{1}, Adrian van Kan\aff{2}, Edgar Knobloch\aff{2}, Keith Julien\aff{1} and Chang Liu\aff{3}}

\affiliation{\aff{1}Department of Applied Mathematics, University of Colorado, Boulder, CO 80309, USA
\aff{2}Department of Physics, University of California at Berkeley, Berkeley, CA 94720, USA
\aff{3}School of Mechanical, Aerospace, and Manufacturing Engineering, University of Connecticut, Storrs, CT 06269, USA}

\corresau{Adrian E. Fraser, \email{adrian.fraser@colorado.edu}}

\begin{document}
\maketitle

\begin{abstract}
We study the dynamics of salt fingers in the regime of slow salinity diffusion (small inverse Lewis number) and strong stratification (large density ratio), focusing on regimes relevant to Earth's oceans. Using three-dimensional direct numerical simulations in periodic domains, we show that salt fingers exhibit rich, multiscale dynamics in this regime, with vertically elongated fingers that are twisted into helical shapes at large scales by mean flows and disrupted at small scales by isotropic eddies. We use a multiscale asymptotic analysis to motivate a reduced set of partial differential equations that filters internal gravity waves and removes inertia from all parts of the momentum equation except for the Reynolds stress that drives the helical mean flow. When simulated numerically, the reduced equations capture the same dynamics and fluxes as the full equations in the appropriate regime. The reduced equations enforce zero helicity in all fluctuations about the mean flow, implying that the symmetry-breaking helical flow is spontaneously generated by strictly non-helical fluctuations.

\end{abstract}

\section{Introduction} \label{sec:intro}
The salt-finger instability occurs in stably stratified fluid layers with background temperature and salinity that both increase with height, and a sufficiently small ratio of salinity diffusivity $\kappa_S$ to thermal diffusivity $\kappa_T$. This instability drives significant turbulent mixing and a broad range of dynamics in the ocean \citep{radko_book}, where this diffusivity ratio---the inverse Lewis number---is quite small: $\tau \equiv \kappa_S/\kappa_T \approx 10^{-2}$.

In stably stratified systems where heat is the sole contributor to buoyancy, large thermal diffusivity has been leveraged to derive asymptotically reduced sets of PDEs valid in the so-called ``low-P\'eclet number" (LPN) limit \citep{Lignieres_LPN,garaud_journey_2021}, where the buoyancy equation reduces to a diagnostic balance between advection of the background temperature gradient and diffusion of thermal fluctuations. Given that rapid thermal diffusion is fundamental to the salt-finger instability, one might naturally expect similar asymptotic reductions to be applicable. Indeed, \citet{Prat_smallPr} explored the LPN limit for salt fingers in astrophysical regimes (cf.~\citealt{Spruit}), where both $\tau$ and the ratio of viscous to thermal diffusion, the Prandtl number $\Pr \equiv \nu/\kappa_T$, are extremely small ($\Pr$, $\tau \sim 10^{-6}$; \citealt{garaud_double-diffusive_2018}). They found that the LPN limit reproduces the same turbulent fluxes as the full equations in the appropriate limit. The \textcolor{black}{low salinity diffusivity limit} was also studied, albeit in 2D, by \citet{xie_reduced_2017}, who showed, in addition, that in the oceanographic regime of $\Pr \gtrsim O(1)$, the momentum equation reduces to a diagnostic balance involving buoyancy and viscosity. In this regime, the evolution is driven by the salinity field alone, with subdominant inertial terms, resulting in inertia-free salt convection (IFSC).

The reductions offered by these limits simplify both numerical and analytical computations while excluding presumably irrelevant dynamics in their respective regimes of validity. For instance, in the LPN limit internal gravity waves are overdamped, and thus a large buoyancy frequency no longer constrains the simulation time step in this limit. However, the regions in parameter space where the excluded dynamics remain important are not always clear \emph{a priori}. The spontaneous formation of thermohaline staircases and the large-scale, secondary instabilities that often precede them (e.g., the collective and gamma instabilities, see \citealt{radko_mechanism_2003,traxler_dynamics_2011}) are excluded in the LPN limit, but these can still occur when $\tau$ and/or $\Pr$ are extremely small, provided the system is not too strongly stratified \citep{garaud_double-diffusive_2018}. Thus, one expects the LPN and IFSC limits to faithfully capture the dynamics of salt fingers provided $\tau$ and/or $\Pr$ are sufficiently small \emph{and} the density stratification is sufficiently large.

With these uncertainties in mind, we extend here the work of \citet{xie_reduced_2017} to three dimensions, performing a suite of direct numerical simulations (DNS) of both the primitive and IFSC equations at $\tau = 0.05$ and $\Pr = 5$ with varying degrees of stratification, focusing on the limit of \textcolor{black}{moderate or} strong stratification (\textcolor{black}{moderate or} weak instability). 
We find that this regime is characterized by remarkably rich, multiscale dynamics\textcolor{black}{, including helical mean flows 
(resembling the mean flows seen in bounded domains by \citealt{yang_scaling_2016})
that are not captured by the IFSC reduction of \citet{xie_reduced_2017}.} 
Motivated by the simulation results, we consider a multiscale asymptotic expansion of our system, which points to a natural modification of the IFSC model. 
This modified IFSC (MIFSC) model reproduces the dynamics seen in simulations of the full equations for much weaker stratification (stronger instability)\textcolor{black}{, provides an explanation and reduced description of the mean 
flows observed by \citet{yang_scaling_2016},} and suggests how the fields and fluxes might scale with stratification, which we show to be broadly consistent with the simulations.

\section{Problem formulation} \label{sec:methods}
We are interested in the dynamics of salt fingers in the simultaneous limits of fast thermal diffusion and weak or moderate instability. We consider fluctuations atop linear background profiles of salinity and potential temperature in the vertical with constant slopes $\beta_S$ and $\beta_T$, respectively. 
We assume the flows are slow enough and the layer height small enough to permit the use of the Boussinesq approximation.
In this limit, the standard control parameters include the Prandtl number $\Pr \equiv \nu/\kappa_T$, the inverse Lewis number $\tau \equiv \kappa_S/\kappa_T$, and the density ratio $R_\rho \equiv \alpha_T \beta_T / (\alpha_S \beta_S)$ with $\alpha_T>0$, $\alpha_S>0$ being the respective coefficients of expansion. 
We consider periodic boundary conditions in all directions, in which case our system is linearly unstable to the salt-finger instability for $1 < R_\rho < \tau^{-1}$ \citep{baines_thermohaline_1969}, with $R_\rho = \tau^{-1}$ corresponding to marginal diffusive stability and $R_\rho < 1$ to an unstably stratified background and hence dynamical instability. In the regime of interest here, it is helpful to introduce the following control parameters:
\begin{equation}
    \Ra \equiv \frac{\alpha_S \beta_S \kappa_T}{\alpha_T \beta_T \kappa_S} = \frac{1}{R_\rho \tau}, \quad \R \equiv \Ra - 1 \quad \text{and} \quad \Sc \equiv \frac{\nu}{\kappa_S} = \frac{\Pr}{\tau},
\end{equation}
where $\Ra$ is the Rayleigh ratio (with marginal stability now at $\Ra = 1$), $\R$ is a measure of supercriticality and $\Sc$ is the Schmidt number. In all results presented below, we fix $\Pr = 5$, $\tau = 0.05$ and thus $\Sc = 100$. 

We follow Sec.~3.1 of \citet{xie_reduced_2017} in our choice of nondimensionalization, taking the characteristic finger width $d = (\kappa_T \nu/g \alpha_T \beta_T)^{1/4}$ (with gravitational acceleration $g$) as the length scale and the salinity diffusion time $d^2/\kappa_S$ as the timescale. As our temperature scale, we take the background temperature difference across a height $d$ rescaled by $\tau$, yielding $\tau \beta_T d$. Similarly, our unit for salinity fluctuations is the background salinity difference across $d$ rescaled by $\Ra^{-1}$, yielding $\Ra^{-1} \beta_S d$. The governing equations\textcolor{black}{, with hats over dependent and independent variables (and derivatives) to indicate this choice of non-dimensionalisation,} are
\begin{equation} \label{eq:full_mom}
    \frac{1}{\Sc} \left( \frac{\partial}{\partial \hat t} + \hat{\mathbf{u}} \cdot \hat \nabla \right) \hat{\mathbf{u}} = - \hat \nabla \hat p + (\hat T - \hat S) \mathbf{e}_z + \hat \nabla^2 \hat{\mathbf{u}},
\end{equation}
\begin{equation} \label{eq:full_cont}
    \hat \nabla \cdot \hat{\mathbf{u}} = 0,
\end{equation}
\begin{equation} \label{eq:full_T}
    \tau \left( \frac{\partial}{\partial \hat t} + \hat{\mathbf{u}} \cdot \hat \nabla \right) \hat T + \hat w = \hat \nabla^2 \hat T,
\end{equation}
and
\begin{equation} \label{eq:full_S}
    \left( \frac{\partial}{\partial \hat t} + \hat{\mathbf{u}} \cdot \hat \nabla \right) \hat S + \Ra\, \hat w = \hat \nabla^2 \hat S,
\end{equation}
with velocity $\hat{\mathbf{u}} = (\hat{u}, \hat{v}, \hat w)$ and temperature, salinity, and pressure fluctuations $\hat T$, $\hat S$ and $\hat p$. \textcolor{black}{Note that $\hat T$ and $\hat S$ represent fluctuations atop the imposed background profiles of potential temperature and salinity, which vary linearly with depth.} \textcolor{black}{These equations are identical to those that have been used to study salt fingers in triply periodic boxes elsewhere \citep[see, e.g.,][]{radko_book,garaud_double-diffusive_2018}, albeit with a different nondimensionalisation, and thus with different control parameters appearing in different places.}


In what follows, we present DNS of the above system for different values of $\Ra$. These simulations were performed using the pseudospectral-tau method implemented in Dedalus v2 \citep{Dedalus2}. We use periodic boundary conditions with a horizontal domain size of $(L_x, L_y) = (4 \times 2\pi/\hat{k}_\mathrm{opt}, 2 \times 2\pi/\hat{k}_\mathrm{opt})$, where $\hat{k}_\mathrm{opt}(\Pr, \tau, R_\rho)$ is the horizontal wavenumber at which the linear instability is most unstable for a given set of parameters (for a subset of our parameters, we have checked that the dynamics and time-averaged fluxes change negligibly upon increasing $L_x$ and $L_y$). For the parameters of interest, the salt fingers become very extended in $z$, and thus our domain height must be very large to avoid artificial domain-size effects \citep[see, e.g., Appendix A.3 of][]{traxler_dynamics_2011}. For most of the cases reported here, we find $L_z = 64 \times 2\pi/\hat{k}_\mathrm{opt}$ to be sufficient (where we confirm this by comparing against results with $L_z = 128 \times 2\pi/\hat{k}_\mathrm{opt}$), although we find that shorter domains suffice for $\R \gtrsim 1$, while taller domains are necessary for $\R \lesssim 1/8$. We dealias using the standard 3/2-rule and use a numerical resolution of 8 Fourier modes per $2 \pi/\hat{k}_\mathrm{opt}$ in each direction,
although twice this resolution becomes necessary (as verified by convergence checks) for our simulations with the largest values of $\R$. Note that, per our dealiasing procedure, nonlinearities are evaluated on a grid with $3/2$ times this resolution. We timestep using a semi-implicit, second-order Adams-Bashforth/backwards difference scheme (Dedalus' ``SBDF2" option, Eq.~[2.8] of \citealt{Wang_timesteppers_2008}), with nonlinearities treated explicitly and all other terms treated implicitly, and an advective CFL safety factor of $0.3$ (sometimes $0.15$ for our highest $\R$ cases). 
We initialize simulations with small-amplitude noise in $S$\textcolor{black}{, and all other fields set to zero}.


\section{Trends across \texorpdfstring{$\R$}{R}} \label{sec:dynamics}
\begin{figure}
    \centering
    \includegraphics[width=\textwidth]{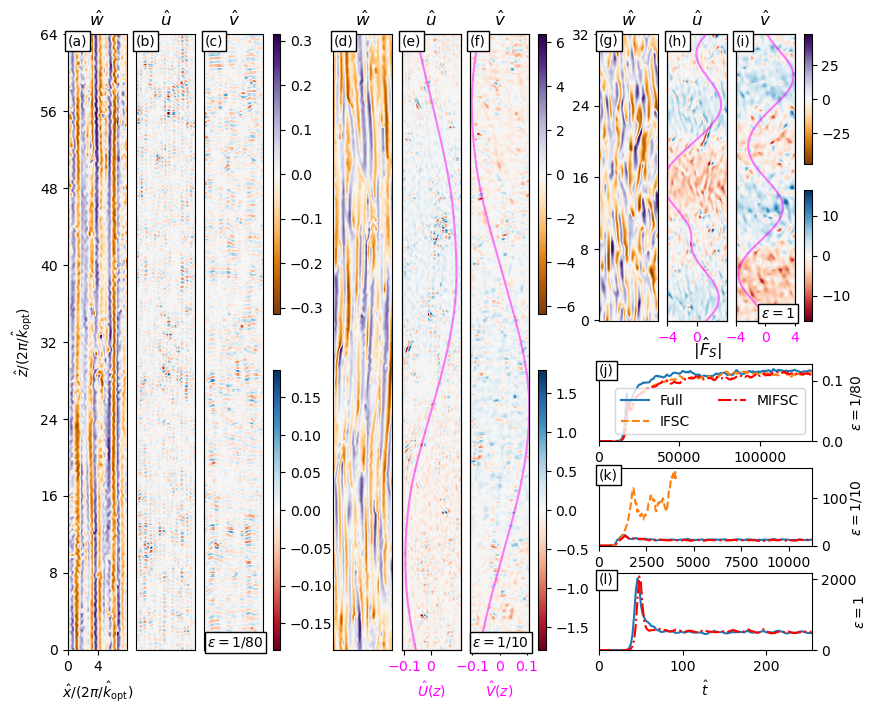}
    \caption{Flow velocity snapshots at $y=0$ in the saturated state from simulations of equations \eqref{eq:full_mom}-\eqref{eq:full_S} with varying supercriticality: $\R = 1/80$ (a-c), $\R = 1/10$ (d-f) and $\R = 1$ (g-i), with timetraces of the corresponding salinity flux $|F_S|$ \textcolor{black}{(blue solid lines)} shown in panels j, k, and l, respectively, alongside fluxes from different reduced models (\textcolor{black}{orange dashed and red dash-dotted lines,} see Sec.~\ref{sec:models}). All cases exhibit a multiscale and anisotropic flow where fingers with large vertical extent and vertical velocity (compared to horizontal width and velocity) coexist with small-scale, isotropic disturbances. Magenta curves (panels e, f, h, and i) show the time-average (over the saturated state) of the horizontal, helical mean flow $\hat{\mathbf{U}}_\perp$ that becomes a significant feature for $\R \gtrsim 0.1$.
    }
    \label{fig:dynamics}
\end{figure}
Figure \ref{fig:dynamics} shows velocity snapshots from simulations with $\R = 1/80$ (panels a-c), $1/10$ (panels d-f) and $1$ (panels g-i), i.e., $R_\rho \simeq 19.75$, $18.18$ and $10$, illustrating general trends in this regime. In each case, we see highly anisotropic and multiscale dynamics, with vertically elongated, large-amplitude structures (the characteristic salt fingers) in $w$, and smaller-amplitude, isotropic eddies seen in each velocity field. The separation between the long vertical and the short isotropic scales shrinks as $\R$ increases.

At very small $\R$ (see panels a-c), the fingers become vertically invariant ``elevator modes"\footnote{At larger $\R$, self-connecting structures only persist for insufficiently tall domains and lead to bursty and domain height-dependent dynamics. At very small $\R$, self-connecting structures persist in even the tallest domains we can reasonably achieve numerically, but they do not drive bursty dynamics or domain height-dependent dynamics.} disturbed by isotropic ripples. 
For moderate supercriticality, $\R \sim 0.1$, the fingers are no longer vertically invariant but still very anisotropic, with much larger vertical scale and velocity than in the horizontal.

\begin{figure}
    \centering
    \includegraphics[width=\textwidth]{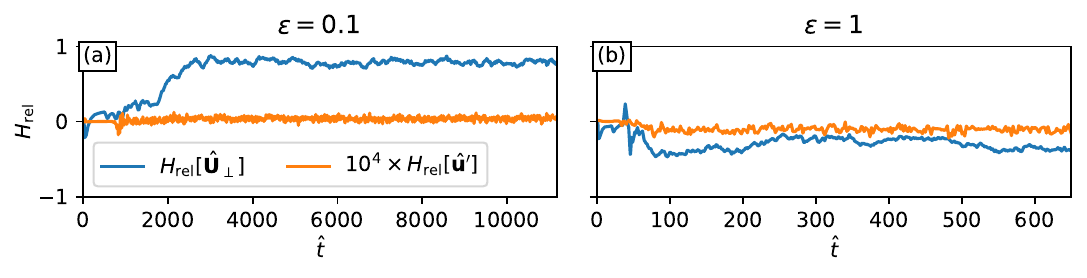}
    \caption{Relative helicity (see text) of the mean flow (blue) and of the fluctuations about the mean flow (orange; multiplied by $10^4$ to ease comparison) for two values of $\R$. At small $\R$, the flow is almost maximally helical, and in both cases the fluctuations are highly non-helical, with $H_\mathrm{rel}[\hat{\mathbf{u}}'] \sim 10^{-5}$.
    }
    \label{fig:helicity}
\end{figure}

In this regime, a horizontal mean flow $\hat{\mathbf{U}}_\perp$ 
develops spontaneously (cf.~\citealt{chang}), as shown by the magenta curves in panels e-f. The two components of $\hat{\mathbf{U}}_\perp$ are $\pi/2$ out of phase in $z$---i.e., one component passes through $0$ as the other reaches an extremum---resulting in a
\textcolor{black}{mean flow with nonzero helicity $H[\hat{\mathbf{U}}_\perp]$, where
\begin{equation}
    H[\hat{\mathbf{u}}] = \int \hat{\mathbf{u}} \cdot \left( \hat \nabla \times \hat{\mathbf{u}} \right) {\rm d}V
\end{equation}
defines the helicity of a given flow field $\hat{\mathbf{u}}$.}
\textcolor{black}{As the mean flow has no vertical component, its helicity arises from the horizontal components of $\hat{\mathbf{U}}_\perp$ and $\hat \nabla \times \hat{\mathbf{U}}_\perp$, and it} advects the fingers into a corkscrew-like shape \textcolor{black}{(the graphical abstract accompanying this article---see Fig.~\ref{fig:graphical_abstract}---shows a volume rendering of the vertical velocity for the same $\Ra = 2$ simulation as shown in Fig.~\ref{fig:dynamics}(g-i) and provides a visualisation of this flow).}
In fact, this mean flow is nearly a \emph{maximally helical} (Beltrami) flow in that its relative helicity 
\textcolor{black}{$H_\mathrm{rel}[\hat{\mathbf{U}}_\perp] \equiv H[\hat{\mathbf{U}}_\perp]/[(\int |\hat{\mathbf{U}}_\perp|^2\, {\rm d}V) (\int |\hat \nabla \times \hat{\mathbf{U}}_\perp |^2 {\rm d}V)]^{1/2}$}, 
calculated using the time- and horizontally-averaged flow, is over 0.99---alternatively, $H_\mathrm{rel}$ of the instantaneous mean flow saturates at roughly $0.8$, see Fig.~\ref{fig:helicity}. While $H_\mathrm{rel} > 0$ for this particular simulation, other random initial conditions lead to $H_\mathrm{rel} < 0$ (not shown) with no clear statistical preference between the two signs based on our limited sample. In stark contrast to the strongly helical mean flow, the relative 
helicity of the fluctuations about the mean \textcolor{black}{($\hat{\mathbf{u}}' \equiv \hat{\mathbf{u}} - \hat{\mathbf{U}}_\perp$)} is roughly $10^{-5}$. This leads to the remarkable observation that the system spontaneously forms a symmetry-breaking, maximally helical flow from nonhelical fluctuations. 
\textcolor{black}{In fact the generation of such mean flows by salt fingers was already seen in vertically bounded domains by \citet{yang_scaling_2016} (see their Fig.~16), albeit without comment (see also the discussion of fingers ``twisting about one another" in \citealt{schmitt_effects_1979}). 
Similar spontaneous mirror symmetry breaking and Beltrami flow formation have also been observed in other systems, including active matter  \citep{slomka2017spontaneous,romeo2024vortex} and magnetohydrodynamic  \citep{agoua2021spontaneous} turbulence, but without such strikingly non-helical fluctuations. }

For yet larger $\R$ (see Fig.~\ref{fig:dynamics} panels g-i), both the mean flow and the fingers become more vigorous, and the anisotropy of the fingers is less extreme, permitting shorter vertical domains. In this regime, the mean flow is still very helical at each $z$, but tends to have a shorter vertical scale and its helicity may change sign with $z$ (see Fig.~\ref{fig:helicity} panel b). \textcolor{black}{The helicity is also less stationary in this regime, and for some initial conditions it is observed to change sign with time, similar to the finite lifetimes of different flow patterns observed in convection \citep[see, e.g.,][]{ahlers_time_1985,winchester_zonal_2021,wang_lifetimes_2023}.}

\begin{figure}
    \centering
    \includegraphics[width=\textwidth]{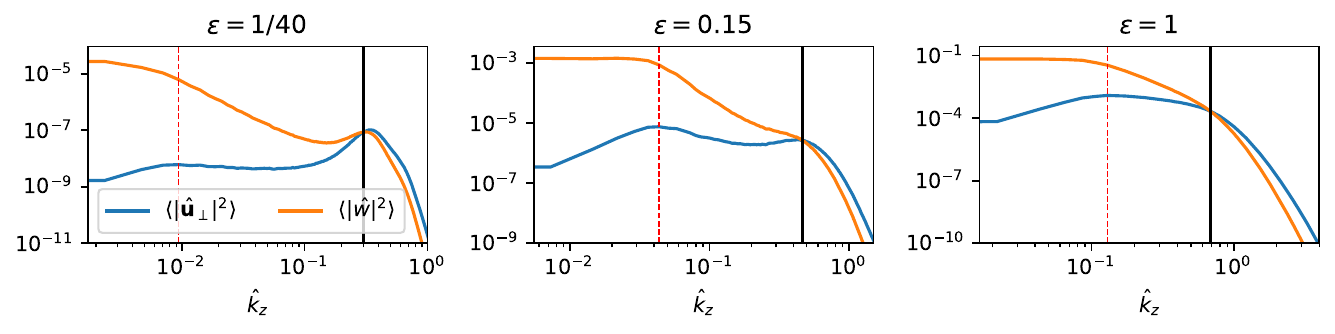}
    \caption{Horizontal (blue) and vertical (orange) kinetic energy spectra \textcolor{black}{(time-averaged over the statistically stationary state)} versus $\hat{k}_z$ at $\hat{k}_y = 0$ and $\hat{k}_x = \hat{k}_\mathrm{opt}$. 
    Black lines show $\hat{k}_z = \hat{k}_\mathrm{opt}$ to highlight the small-scale isotropic flow component while the red vertical lines correspond to  the secondary peak in the horizontal spectrum to highlight the anisotropic, small $\hat{k}_z$ flow component. The ratio between these two wavenumbers provides one measure of anisotropy shown in Fig.~\ref{fig:RMS_trends}.
    }
    \label{fig:kz_spectra}
\end{figure}

Both the multiscale aspect of this system and the trends in scale separation with $\R$ are readily seen in Fig.~\ref{fig:kz_spectra}, which shows spectra of the kinetic energy as a function of the vertical wavenumber $\hat{k}_z$ 
at $\hat{k}_\perp = \hat{k}_\mathrm{opt}$, the fastest-growing wavenumber of the linear instability. Two distinguishing features are seen most clearly in the horizontal kinetic energy, which has a local maximum (or otherwise a clear change in the spectrum, in the case of large $\R$) at isotropic scales where $\hat{k}_z\sim \hat{k}_\mathrm{opt}$, indicated by the black vertical lines, and a local maximum at smaller $\hat{k}_z$ indicated by the red vertical lines. 
Note that the gap in $\hat{k}_z$ between these two local maxima shrinks as $\R$ increases, consistent with the observed decrease in scale separation with increasing $\R$ (Fig.~\ref{fig:dynamics}).

\section{Asymptotic models} \label{sec:models}

\begin{figure}
    \centering
    \includegraphics[width=\textwidth]{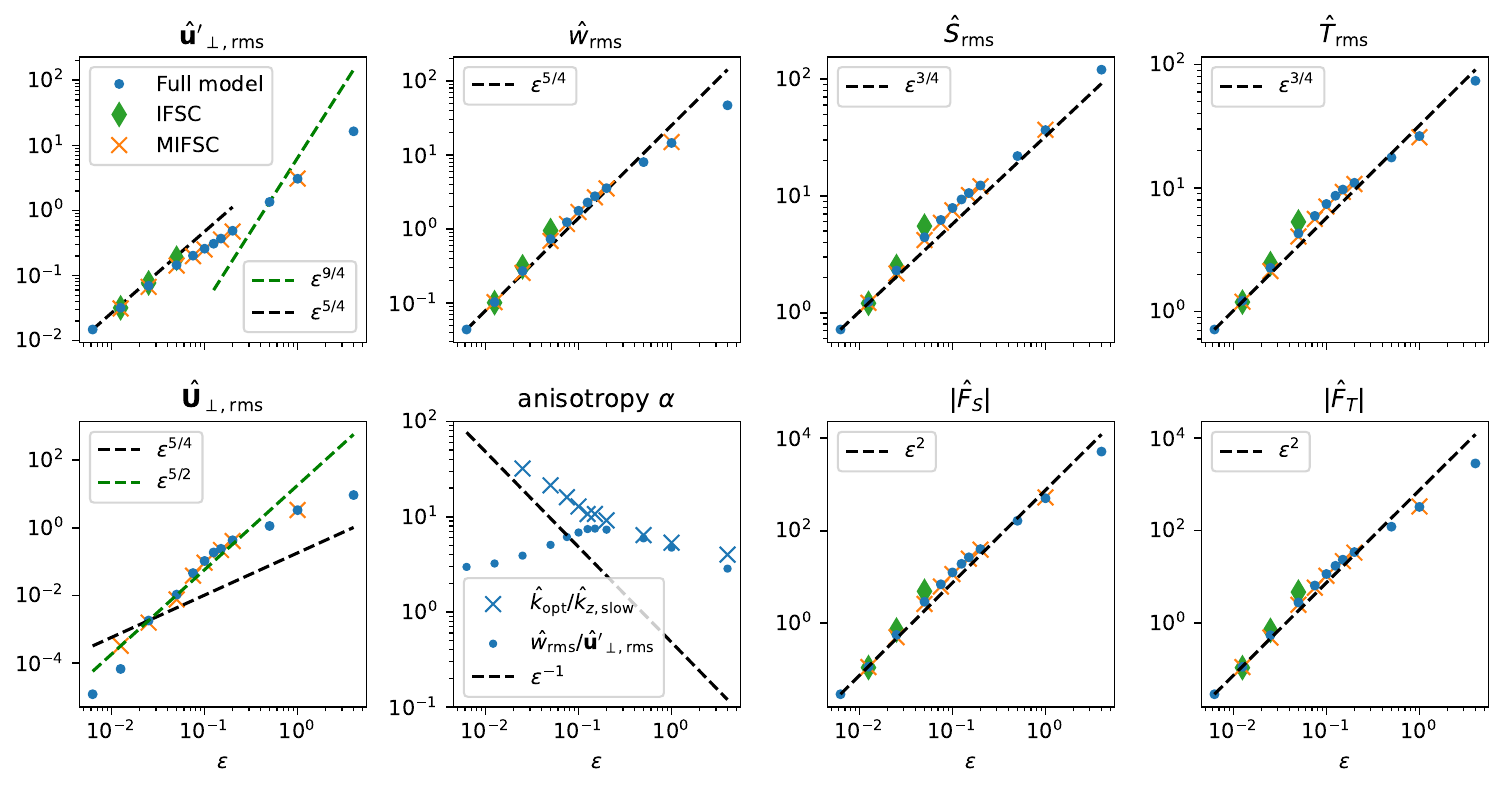}
    \caption{Scalings with respect to $\R$ of several quantities (indicated in the caption for each panel) for the full system, Eqs.~\eqref{eq:full_mom}-\eqref{eq:full_S} (blue dots), the IFSC model, with Eqs.~\eqref{eq:full_T} and \eqref{eq:full_mom} replaced by Eqs.~\eqref{eq:LPN_T} and \eqref{eq:IFSC_mom} (green diamonds), and the modified IFSC (MIFSC) model, where Eq.~\eqref{eq:full_mom} is replaced instead by Eqs.~\eqref{eq:MIFSC_mean_mom}-\eqref{eq:MIFSC_fluct_mom} (orange crosses). Black dashed lines show scalings predicted by the multiscale asymptotic analysis described in the text. The green dashed lines and the two measures of anisotropy are described in the text.
    }
    \label{fig:RMS_trends}
\end{figure}

The IFSC model of \citet{xie_reduced_2017}, in three dimensions (3D), is appropriate when $\tau \ll 1$ and $Sc \gg 1$ and is described by the equations
\begin{equation} \label{eq:LPN_T}
    \hat w = \hat \nabla^2 \hat T,
\end{equation}
\begin{equation} \label{eq:IFSC_mom}
    0 = - \hat \nabla \hat p + (\hat T - \hat S) \mathbf{e}_z + \hat \nabla^2 \hat{\mathbf{u}},
\end{equation}
with Eqs.~\eqref{eq:full_cont} and \eqref{eq:full_S} left unchanged. While it appears that the model should be valid for all order-one $\R$, we show in Fig.~\ref{fig:dynamics} that this is not the case: the model does produce dynamics and fluxes consistent with the full system at sufficiently small supercriticality, roughly $\R \lesssim 1/80$, but for $\R \approx 1/20$ or larger, it produces dynamics that differ qualitatively from solutions of the full equations---while the full system exhibits helical mean flows that disrupt and twist the fingers on large scales, the IFSC model removes the Reynolds stress term from the horizontal mean of Eq.~\eqref{eq:full_mom} and thus lacks these flows. Without mean flows to disrupt the long fingers, such fingers drive temporally bursty dynamics that dramatically raise the fluxes, as shown by the IFSC curve in panel k of Fig.~\ref{fig:dynamics}. 

However, the IFSC model can be modified to capture mean flow generation. Our simulations suggest that anisotropy is an essential aspect of a reduced description of salt fingers valid in the regime depicted in Fig.~\ref{fig:dynamics}, possibly with a rescaling of the various fields to retain the Reynolds stress at leading order. In order to capture both the elongated fingers shown in Fig.~\ref{fig:dynamics} and the small scale isotropic fluctuations therein, we employ a multiscale asymptotic analysis inspired by a related approach to turbulence in stably stratified fluids employed by \citet{chini_exploiting_2022},  \citet{shah_regimes_2024}, and \citet{garaud_numerical_2024}.

To begin, we note that the growth rate and optimal wavenumber of the linear instability scale as $\R^{3/2}$ and $\R^{1/4}$, respectively, for sufficiently small $\R$---for the $\Pr$, $\tau$ considered here, this scaling is achieved for $\R \lesssim 1$ (see, e.g., Fig.~3 of \citealt{xie_reduced_2017}). It is thus convenient to rescale the independent and dependent variables as follows (cf.~\citealt{radko_equilibration_2010}):
\begin{equation}
    \hat{\mathbf{x}} = \R^{-1/4} \mathbf{x}, \quad \hat t = \R^{-3/2} t, \quad \hat{\mathbf{u}} = \R^{5/4} \mathbf{u}, \quad \hat p = \R^{3/2} p, \quad (\hat T, \hat S) = \R^{3/4} (T,S).
\end{equation}
Equations \eqref{eq:full_mom}-\eqref{eq:full_S} then become
\begin{equation} \label{eq:small-R_mom}
    \R \frac{1}{\Sc} \left( \frac{\partial}{\partial t} + \mathbf{u} \cdot \nabla \right) \mathbf{u} = - \nabla p + \R^{-1} (T - S) \mathbf{e}_z + \nabla^2 \mathbf{u},
\end{equation}
\begin{equation} \label{eq:small-R_T}
    \R \tau \left( \frac{\partial}{\partial t} + \mathbf{u} \cdot \nabla \right) T + w = \nabla^2 T,
\end{equation}
and
\begin{equation} \label{eq:small-R_S}
    \R \left( \frac{\partial}{\partial t} + \mathbf{u} \cdot \nabla \right) S + \Ra\, w = \nabla^2 S,
\end{equation}
with Eq.~\eqref{eq:full_cont} left unchanged. 
When $\R={\cal O}(1)$ but $\tau\ll1$ and $Sc\gg1$, the inertial terms in Eqs.~(\ref{eq:small-R_mom})-(\ref{eq:small-R_T}) drop out and the resulting equations correspond to the 3D IFSC model with Eq.~(\ref{eq:small-R_S}) providing the sole prognostic equation. 
On the other hand, when $\R\ll1$, we may expand all fields as asymptotic series in $\R$ as $q = \sum_n q_n \R^n$. Inspecting the $z$ component of Eq.~\eqref{eq:small-R_mom} in the limit $\R\to 0$ shows that $T_0 = S_0$, i.e., the dynamics are neutrally buoyant (or asymptotically spicy) in this limit. In the following it is helpful to introduce the buoyancy $b \equiv \R^{-1}(T-S)$ and subtract Eq.~\eqref{eq:small-R_T} from Eq.~\eqref{eq:small-R_S}, yielding
\begin{equation} \label{eq:small-R_b}
    \left( \frac{\partial}{\partial t} + \mathbf{u} \cdot \nabla \right) (S - \tau T) + w = -\nabla^2 b.
\end{equation}

\textcolor{black}{We now perform a multi-scale asymptotic expansion employing the following four steps:
\begin{enumerate}
    \item Introduce fast and slow scales in $z$ and $t$ and allow each field $q$ to depend on them both, i.e., $q = q(\mathbf{x}_\perp, z_s, z_f, t_s, t_f)$, with $\partial_z \mapsto \alpha^{-1} \partial_{z_s} + \partial_{z_f}$ and $\partial_t \mapsto \alpha^{-1} \partial_{t_s} + \partial_{t_f}$; 
    here the fast vertical scale $z_f$ is isotropic, and the anisotropy of the slow vertical scale $z_s$ is controlled by $\alpha$, which we take to be $\alpha = \Sc/\R \gg 1$ (which we note is qualitatively consistent with the trends seen in Figs.~\ref{fig:dynamics} and \ref{fig:kz_spectra}).
    \item Decompose each field $q$ into its fast-averaged and fluctuating components: 
    $q = \fslow{q} + \ffast{q}$ where $\fslow{\cdot}$ denotes an average over $z_f$ and $t_f$, so that $\fslow{q}$ depends only on $\mathbf{x}_\perp$, $z_s$, and $t_s$, while $\ffast{q}$ depends on all coordinates. The only exception to this $q = \fslow{q} + \ffast{q}$ decomposition is for the pressure and horizontal flow fluctuations (defined as $\mathbf{u}'_\perp \equiv \mathbf{u}_\perp - \mathbf{U}_\perp$), where we instead take $p = \alpha^{-1} \fslow{p} + \ffast{p}$ and $\mathbf{u}'_\perp = \alpha^{-1} \fslow{\mathbf{u}'_\perp} + \ffast{\mathbf{u}}'_\perp$.\footnote{\textcolor{black}{Note that this is without loss of generality: we are simply taking different non-dimensionalisations for $\fslow{p}$ and $\ffast{p}$ and likewise for $\fslow{\mathbf{u}'_\perp}$ and $\ffast{\mathbf{u}}'_\perp$.}}
    \item Take the fast average of each of the governing equations to obtain an equation for $\partial_{t_s} \fslow{q}$, and subtract the result from the original equation to obtain an equation for $\partial_t \ffast{q}$.
    \item Finally, retain only leading-order terms by removing terms that are small in the limits $\R \to 0$, $\tau \to 0$, $\Sc \to \infty$, and $\alpha \to \infty$ (with $\alpha \R/\Sc = 1$).
\end{enumerate}
An explicit demonstration of these steps is provided in, e.g., \cite{chini_exploiting_2022} and \cite{shah_regimes_2024}.}

\textcolor{black}{The resulting equations are:}
\begin{equation}
    \fslow{w} = \nabla_\perp^2 \fslow{T} \quad \text{and} \quad \ffast{w} = \nabla^2_f \ffast{T},
\end{equation}
\begin{equation}
\fslow{\mathbf{U}_\perp} \cdot \nabla_\perp \fslow{S} + \fslow{\ffast{\mathbf{u}} \cdot \nabla_f \ffast{S}} + \fslow{w} = - \nabla_\perp^2 \fslow{b},
\end{equation}
\begin{multline}
\left[ \frac{\partial}{\partial t_f} + \fslow{\mathbf{U}_\perp} \cdot \nabla_\perp + \fslow{w} \frac{\partial}{\partial z_f} \right] \ffast{S} + \ffast{\mathbf{u}} \cdot \nabla_f \ffast{S} - \fslow{\ffast{\mathbf{u}} \cdot \nabla_f \ffast{S}} \\ 
+ \ffast{\mathbf{u}}_\perp \cdot \nabla_\perp \fslow{S} + \ffast{w} = - \nabla_f^2 \ffast{b},
\end{multline}
\begin{equation}
    0 = - \nabla_\perp \fslow{p} + \fslow{b} \mathbf{e}_z + \nabla_\perp^2 \fslow{\mathbf{u}'} \quad \text{and} \quad 0 = - \nabla_f \ffast{p} + \ffast{b} \mathbf{e}_z + \nabla_f^2 \ffast{\mathbf{u}}',
\end{equation}
\begin{equation} \label{eq:multiscale_ubar}
    \frac{\partial}{\partial t_s} \fslow{\mathbf{U}_\perp} + \frac{\partial}{\partial z_s}\langle \ffast{\mathbf{u}}_\perp \ffast{w}\rangle_{f,\perp}  = \frac{\partial^2}{\partial z_s^2} \fslow{\mathbf{U}_\perp},
\end{equation}
\begin{equation}
    \nabla_s \cdot \fslow{\mathbf{u}'} = 0 \quad \text{and} \quad \nabla_f \cdot \ffast{\mathbf{u}}' = 0,
\end{equation}
where we have introduced $\nabla_s \equiv (\partial_x, \partial_y, \partial_{z_s})$ and $\nabla_f \equiv (\partial_x, \partial_y, \partial_{z_f})$\textcolor{black}{, with $\langle \cdot \rangle_{f, \perp}$ denoting an average over $\mathbf{x}_\perp$ in addition to $z_f$ and $t_f$}. 
The structure of this system of equations is broadly similar to the IFSC model: the temperature equations reduce to a diagnostic balance between diffusion and advection of the background, the salinity equations retain nonlinearity on fast scales (and so do not yield a quasilinear structure as in the work of \citealt{chini_exploiting_2022}), and the momentum equations for fluctuations about the mean flow involve a dominant balance between pressure, buoyancy, and viscosity. However, in contrast to the IFSC model, Eq.~\eqref{eq:multiscale_ubar} retains
the Reynolds stress term absent from Eq.~\eqref{eq:IFSC_mom}. Thus, modifying the IFSC model to capture the leading-order dynamics of the full equations in this limit merely requires retaining the Reynolds stress in the $\mathbf{k}_\perp = 0$ component of the momentum equation: 
\begin{equation} \label{eq:MIFSC_mean_mom}
    \frac{1}{\Sc} \left(\frac{\partial}{\partial \hat t} \hat{\mathbf{U}}_\perp + \frac{\partial}{\partial \hat z} \langle \hat{\mathbf{u}}'_\perp \hat{w}'\rangle_\perp \right) = \frac{\partial^2}{\partial \hat z^2} \hat{\mathbf{U}}_\perp,
\end{equation}
\begin{equation} \label{eq:MIFSC_fluct_mom}
    0 = - \hat \nabla \hat p' + (\hat T' - \hat S') \mathbf{e}_z + \hat \nabla^2 \hat{\mathbf{u}}',
\end{equation}
with temperature given by Eq.~\eqref{eq:LPN_T}, and Eqs.~\eqref{eq:full_cont} and \eqref{eq:full_S} retained in full.
Figure~\ref{fig:dynamics} shows that 
this modified IFSC (MIFSC) model captures the same dynamics as the full equations even for $\R = O(1)$.

The rescaling applied to arrive at the above multiscale asymptotic system offers a natural suggestion for the scaling of the various fields in this limit. For Eqs.~\eqref{eq:full_mom}-\eqref{eq:full_S}, this analysis predicts that both $\hat S$ and $\hat T$ should scale as $\R^{3/4}$, $\hat{\mathbf{U}}_\perp$ and $\hat w$ as $\R^{5/4}$, and the scaling of $\hat{\mathbf{u}}'_\perp$ should differ between fast and slow scales in $\hat z$, with $\fslow{\mathbf{u}'_\perp} \sim \R^{9/4}$ and $\ffast{\mathbf{u}}'_\perp \sim \R^{5/4}$. The predicted stronger $\R$ dependence of $\fslow{\mathbf{u}'_\perp}$ is consistent with Fig.~\ref{fig:kz_spectra}, which shows that the horizontal kinetic energy peaks at large $\hat{k}_z$ for small $\R$ but at small $\hat{k}_z$ for large $\R$.


We present a more quantitative comparison between these predictions and DNS in Fig.~\ref{fig:RMS_trends} by calculating the root-mean-square (r.m.s.) of $\hat{\mathbf{u}}'_\perp$, $\hat{\mathbf{U}}_\perp$, $\hat w$, $\hat S$ and $\hat T$, and the volume-averaged fluxes $\hat F_S = \langle \hat w \hat S \rangle$ and $\hat F_T = \langle \hat w \hat T \rangle$, with the blue dots corresponding to DNS of the full equations, green diamonds to the IFSC model, and orange crosses to the 
\textcolor{black}{MIFSC} model. Anisotropy is quantified by two means: using the wavenumber ratio corresponding to the two spectral peaks in Fig.~\ref{fig:kz_spectra} 
(crosses; the lowest $\R$ values are suppressed because the low-$\hat{k}_z$ peak is difficult to identify and is likely constrained by domain size), and the ratio $\hat w_\mathrm{rms}/\hat{\mathbf{u}}'_{\perp, \mathrm{rms}}$ (dots). In each panel, black dashed lines correspond to the predicted scalings (for $\hat{\mathbf{u}}'_{\perp, \mathrm{rms}}$, the black line shows the predicted $\ffast{\mathbf{u}}'_\perp \sim \R^{5/4}$ scaling while the green line shows the  $\fslow{\mathbf{u}'_\perp} \sim \R^{9/4}$ scaling). The scalings of most of these quantities are consistent with predictions as $\R \to 0$, especially $\hat w_\mathrm{rms}$, $\hat S_\mathrm{rms}$, $\hat T_\mathrm{rms}$, $\hat F_S$ and $\hat F_T$, with the anisotropy measurement 
inconclusive. In contrast, the r.m.s.~of the mean flow, $\hat{\mathbf{U}}_{\perp, \mathrm{rms}}$, follows the predicted $\R^{5/4}$ scaling at larger $\R$ only; \textcolor{black}{at small $\R$, its scaling is perhaps more consistent with $\R^{5/2}$ (green line), although it is difficult to measure $\hat{\mathbf{U}}_{\perp, \mathrm{rms}}$ accurately at the smallest $\R$ values as the mean flow is then very weak and evolves very slowly.}

\section{Conclusions} \label{sec:conclusions}
We have explored the dynamics of salt fingers in 3D and in the limit of slow salinity diffusion ($\tau \ll 1$, $\Pr/\tau \gg 1$) and weak or moderate instability ($\R \equiv R_\rho^{-1} \tau^{-1} - 1 \leq 4$). This regime was studied in 2D by \citet{xie_reduced_2017}, who showed that the temperature equation reduces to a diagnostic balance akin to the low-P\'eclet number limit of \citet{Lignieres_LPN} (see also \citealt{Prat_smallPr,garaud_journey_2021}) and that the vorticity equation is governed by a diagnostic balance between buoyancy and viscous diffusion; Xie {\rm et al.}~referred to these reduced equations as the inertia-free salt convection (IFSC) model. 

Simulations of the full equations in 3D exhibit significant departures from the 2D system. First, the 3D case is characterized by rich, multiscale dynamics where anisotropic fingers are twisted at large scales by a helical mean flow and disrupted at small scales by isotropic eddies. This helical mean flow is a maximally helical, Beltrami flow in some regimes, with helicity of either sign depending on the random initial conditions,
and provides a striking example of spontaneous symmetry breaking, much as occurs in the systems studied by \cite{slomka2017spontaneous}, \cite{agoua2021spontaneous} and \cite{romeo2024vortex}. Second, 3D simulations of the IFSC model exhibit qualitatively different dynamics and significantly enhanced fluxes over the full equations unless the instability is very weak ($\R \ll 0.1$)---a consequence of the exclusion of Reynolds stresses in the IFSC model.

The observed multiscale dynamics inform a multiscale asymptotic expansion in the supercriticality $\R$ 
where the leading-order equations form a closed system. This analysis identifies the leading-order terms missing from the IFSC model---the Reynolds stresses in the equations for the horizontal mean flow---and predicts the scaling of the various fields and fluxes with $\R$.
Simulations of the full equations 
are consistent with these predictions except for the scaling of the mean flow, which has a stronger dependence on $\R$ than suggested by the asymptotic analysis. Furthermore, simulations of the MIFSC equations---the IFSC equations with Reynolds stresses retained---yield qualitative and quantitative agreement with DNS of the full equations up to $\R \approx 1$, further supporting the derived leading-order balances \textcolor{black}{and shedding light on the physical processes contributing to the observed helical flow: for example, because the reduced equations preclude vertical vorticity, internal gravity waves, and modifications to the mean temperature profile, we know such effects are not necessary to produce our helical flows}.

A noteworthy feature of the MIFSC model 
is that, to leading order, the fluctuations $\hat{\mathbf{u}}'_\perp \equiv \hat{\mathbf{u}}_\perp - \hat{\mathbf{U}}_\perp$ have strictly zero helicity. Thus, the helical flow represents a spontaneous symmetry-breaking instability arising from \emph{asymptotically non-helical} fluctuations, analogous to the development of unidirectional traveling waves in reflection-symmetric systems \citep{knobloch1986doubly}.

Our computations of the r.m.s.~values of the various fields broadly support the scalings with $\R$ predicted from the asymptotic analysis. For comparison, \citet{garaud_numerical_2024} devised a means to extract the scalings of fast-averaged quantities $\fslow{q}$ and their fluctuations $\ffast{q}$ separately, demonstrating that their scalings differed in their system. In the present case, our asymptotic analysis points to fields other than $\hat{\mathbf{u}}'_\perp$ and $p$ exhibiting identical scalings on fast and slow scales. While we find no clear discrepancies in our simulations, future work should extend the approach of \citet{garaud_numerical_2024} to the present system to test these predictions more carefully.

Our reduced equations---and those of \citet{xie_reduced_2017} and \citet{Prat_smallPr}---indicate that for $\tau\ll1$ the dynamics no longer depend on $\Pr$ and $\tau$ separately, and only depend on the combination $\Sc \equiv \Pr/\tau$. Thus, while we have only simulated the full equations at $\Pr = 5$, we may expect our simulations of the reduced equations at $\Sc = 100$ to be consistent with the full system in the astrophysically relevant regime of small $\Pr$, $\tau \sim 10^{-7}$-$10^{-4}$ \citep{Prat_smallPr,fraser_characterizing_2022}.
We are thus led to expect that helical flows may form in the interiors of stars, an exciting prospect due to their tendency to support dynamo growth \citep{rincon_dynamo_2019,tobias_turbulent_2021}.

Both the reduced and the full equations admit (unstable) single-mode solutions that may provide a useful proxy for flux computations in the strongly nonlinear regime, and investigations of the role played by the helical mean flow. Outstanding questions involve possible reversals of this flow in longer simulations, and the generation of such flows in vertically confined domains. The multiparameter nature of the problem raises additional questions involving distinct asymptotic regimes when both $\tau$ and $\Sc^{-1}$ are small and if $\Sc=O(1)$ in the regime of small $\tau$ and $\Pr$---note that $\Sc \sim 10^2$ is likely in stars \citep{Skoutnev_zones_of_TI}.

\begin{bmhead}[Acknowledgements]
K.J.~passed away before this manuscript was finalised. We have attempted to present
the results of our collaboration in accordance with his high standards. Any errors or misinterpretations remain our own. 
A.F.~thanks Ian Grooms and Brad Hindman for many useful discussions about helicity and multiscale asymptotic analyses; E.K.~acknowledges an insightful discussion with Nobumitsu Yokoi on the topic of helicity generation in turbulent flow.
\end{bmhead}


\begin{bmhead}[Funding]
This work was supported by NSF under grants OCE-1912332 (A.F.~and K.J.), AST-2402142 (A.F.) and DMS-2308337 (A.v.K.~and E.K.). The computational resources were provided by the NSF ACCESS program (projects PHY250118 and PHY230056), allowing us to utilise the Purdue Anvil CPU cluster. 
\end{bmhead}

\begin{bmhead}[Declaration of interests]
The authors report no conflict of interest.
\end{bmhead}

\appendix
\section{Graphical Abstract and Supplemental Materials}
Here, we include various materials that are associated with the version of this manuscript that was accepted for publication but do not appear in the main text. For completeness, we instead include them here in the arXiv version as appendices. Namely, they are the Graphical Abstract (see Fig.~\ref{fig:graphical_abstract}) and the three sections of supplemental materials, below.

\begin{figure}
    \centering
    \includegraphics[width=0.9\textwidth]{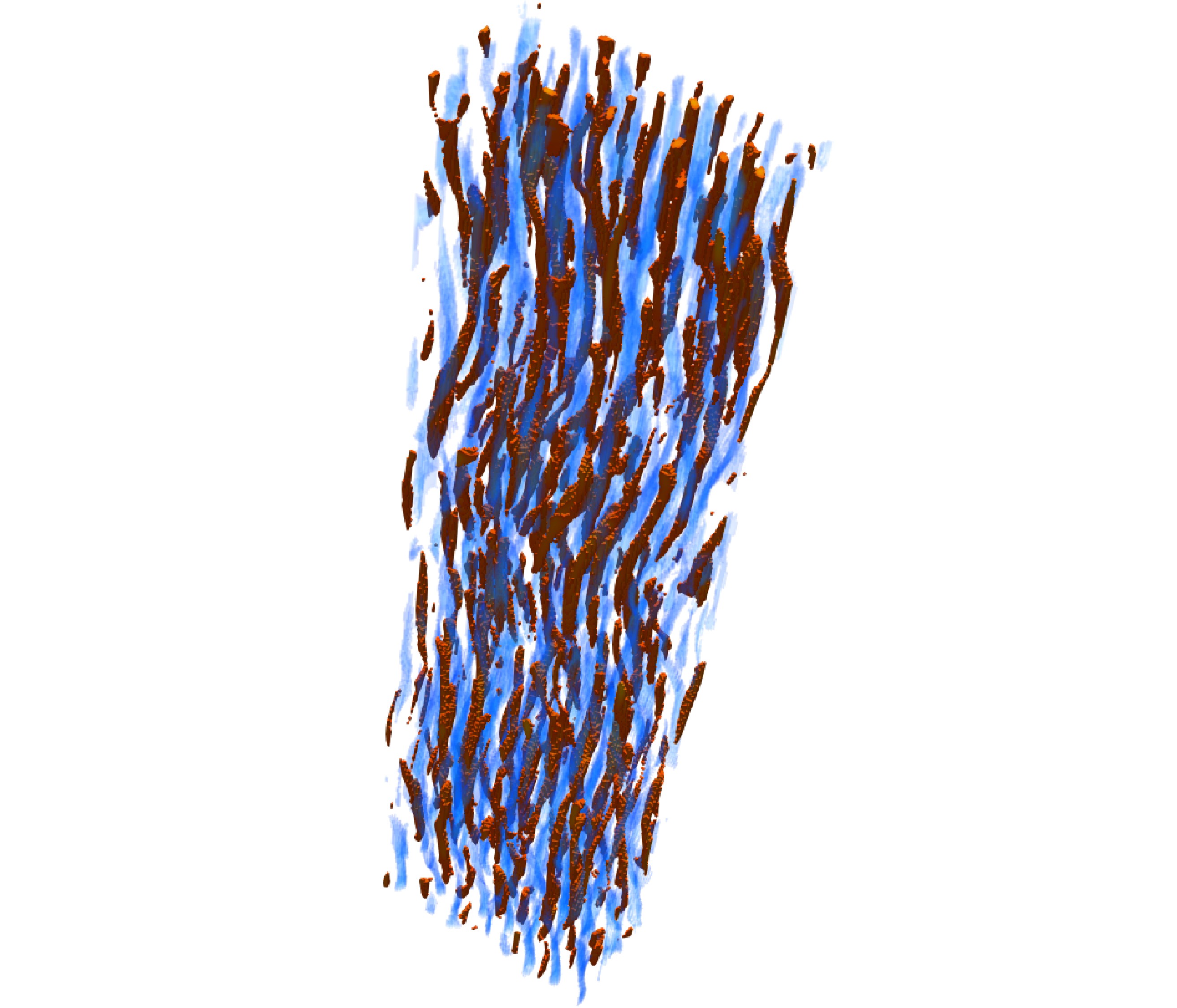}
    \caption{Volume rendering of the vertical velocity for the same $\Ra = 2$ simulation as shown in Fig.~\ref{fig:dynamics}(g-i).
    }
    \label{fig:graphical_abstract}
\end{figure}

\section{Fast-average and fluctuations for full equations} \label{sec:supp:full_eqns}
Start from the isotropic single-scale $\R \to 0$ equations [Eqns.~(4.4)-(4.7) and (2.3) of the main text], introduce $\partial_z \mapsto \partial_{z_f} + \alpha^{-1} \partial_{z_s}$ and $\partial_t \mapsto \partial_{t_f} + \alpha^{-1} \partial_{t_s}$, and rescale $\fslow{\mathbf{u}'_\perp} \mapsto \alpha^{-1} \fslow{\mathbf{u}'_\perp}$ and $\fslow{p} \mapsto \alpha^{-1} \fslow{p}$. Then, break each dynamical equation into its fast-averaged component and the fluctuating component (by subtracting the fast-averaged component from the full equation). In what follows, we present the results of this decomposition, and highlight terms in blue that are subdominant in the limit $\tau \ll 1$, $\Sc \gg 1$, $\R \ll 1$, and $\alpha \gg 1$. This is equivalent to following steps 1-3 (but not 4) shortly after Eq.~(4.7) in the main text.

\subsection{Salinity}
The fast average of the salinity equation is
\begin{equation}
\begin{split}
    & \textcolor{blue}{\R \left[ \frac{1}{\alpha} \frac{\partial}{\partial t_s} + (\fslow{\mathbf{U}_\perp} + \frac{1}{\alpha} \fslow{\mathbf{u'}_\perp}) \cdot \nabla_\perp + \frac{1}{\alpha} \fslow{w} \frac{\partial}{\partial z_s} \right] \fslow{S}} \\
    & \textcolor{blue}{+ \R \fslow{\left[ \ffast{\mathbf{u}}_\perp \cdot \nabla_\perp + \ffast{w} \left( \frac{1}{\alpha} \frac{\partial}{\partial z_s} + \frac{\partial}{\partial z_f} \right) \right] \ffast{S}}} + \Ra \fslow{w} = \left( \nabla_\perp^2 + \textcolor{blue}{\frac{1}{\alpha^2} \frac{\partial^2}{\partial z_s^2}} \right) \fslow{S}.
\end{split}
\end{equation}

Subtracting the fast-average from the full salinity equation then yields the fluctuating part, 
\begin{equation}
\begin{split}
    & \textcolor{blue}{\R \left[ \frac{1}{\alpha} \frac{\partial}{\partial t_s} + \frac{\partial}{\partial t_f} + (\fslow{\mathbf{U}_\perp} + \frac{1}{\alpha} \fslow{\mathbf{u'}_\perp}) \cdot \nabla_\perp + \fslow{w} \left( \frac{1}{\alpha} \frac{\partial}{\partial z_s} + \frac{\partial}{\partial z_f} \right) \right] \ffast{S}} \\
    & \quad \textcolor{blue}{+ \R \left[ \ffast{\mathbf{u}}_\perp \cdot \nabla_\perp + \ffast{w} \left( \frac{1}{\alpha} \frac{\partial}{\partial z_s} + \frac{\partial}{\partial z_f} \right) \right] \ffast{S} - \R \fslow{\left[ \ffast{\mathbf{u}}_\perp \cdot \nabla_\perp + \ffast{w} \left( \frac{1}{\alpha} \frac{\partial}{\partial z_s} + \frac{\partial}{\partial z_f} \right) \right] \ffast{S}}} \\
    & \quad \textcolor{blue}{+ \R \left[ \ffast{\mathbf{u}}_\perp \cdot \nabla_\perp + \ffast{w} \frac{1}{\alpha} \frac{\partial}{\partial z_s} \right] \fslow{S}} + \Ra \ffast{w} \\
    & \qquad = \left( \nabla_\perp^2 + \textcolor{blue}{\frac{1}{\alpha^2} \frac{\partial^2}{\partial z_s^2} + \frac{2}{\alpha} \frac{\partial^2}{\partial z_s \partial z_f}} + \frac{\partial^2}{\partial z_f^2} \right) \ffast{S}.
\end{split}
\end{equation}

\subsection{Temperature}
The temperature equations are similarly
\begin{equation}
\begin{split}
    & \textcolor{blue}{\tau \fslow{\left[ \ffast{\mathbf{u}}_\perp \cdot \nabla_\perp + \ffast{w} \left( \frac{1}{\alpha} \frac{\partial}{\partial z_s} + \frac{\partial}{\partial z_f} \right) \right] \ffast{T}}} \\
    & \quad \textcolor{blue}{+ \tau \left[ \frac{1}{\alpha} \frac{\partial}{\partial t_s} + (\fslow{\mathbf{U}_\perp} + \frac{1}{\alpha} \fslow{\mathbf{u'}_\perp}) \cdot \nabla_\perp + \frac{1}{\alpha} \fslow{w} \frac{\partial}{\partial z_s} \right] \fslow{T}} + \R^{-1} \fslow{w} \\
    & \qquad = \R^{-1} \left( \nabla_\perp^2 + \frac{1}{\alpha^2} \frac{\partial^2}{\partial z_s^2} \right) \fslow{T},
\end{split}
\end{equation}
and
\begin{equation}
\begin{split}
    & \textcolor{blue}{\tau \left[ \frac{1}{\alpha} \frac{\partial}{\partial t_s} + \frac{\partial}{\partial t_f} + (\fslow{\mathbf{U}_\perp} + \frac{1}{\alpha} \fslow{\mathbf{u'}_\perp}) \cdot \nabla_\perp + \fslow{w} \left( \frac{1}{\alpha} \frac{\partial}{\partial z_s} + \frac{\partial}{\partial z_f} \right) \right] \ffast{T}} \\
    & \quad \textcolor{blue}{+ \tau \left[ \ffast{\mathbf{u}}_\perp \cdot \nabla_\perp + \ffast{w} \left( \frac{1}{\alpha} \frac{\partial}{\partial z_s} + \frac{\partial}{\partial z_f} \right) \right] \ffast{T} - \tau \fslow{\left[ \ffast{\mathbf{u}}_\perp \cdot \nabla_\perp + \ffast{w} \left( \frac{1}{\alpha} \frac{\partial}{\partial z_s} + \frac{\partial}{\partial z_f} \right) \right] \ffast{T}}} \\
    & \quad \textcolor{blue}{+ \tau \left[ \ffast{\mathbf{u}}_\perp \cdot \nabla_\perp + \ffast{w} \frac{1}{\alpha} \frac{\partial}{\partial z_s} \right] \fslow{T}} + \R^{-1} \ffast{w} \\
    & \qquad = \R^{-1} \left( \nabla_\perp^2 + \textcolor{blue}{\frac{1}{\alpha^2} \frac{\partial^2}{\partial z_s^2} + \frac{2}{\alpha} \frac{\partial^2}{\partial z_s \partial z_f}} + \frac{\partial^2}{\partial z_f^2} \right) \ffast{T}.
\end{split}
\end{equation}

\subsection{Buoyancy}
The difference of the two yields
\begin{equation}
\begin{split}
    & \fslow{\left[ \ffast{\mathbf{u}}_\perp \cdot \nabla_\perp + \ffast{w} \left( \textcolor{blue}{\frac{1}{\alpha} \frac{\partial}{\partial z_s}} + \frac{\partial}{\partial z_f} \right) \right] (\ffast{S} \textcolor{blue}{- \tau\ffast{T}})} \\
    & \quad + \left[ \frac{1}{\alpha} \frac{\partial}{\partial t_s} + (\fslow{\mathbf{U}_\perp} + \textcolor{blue}{\frac{1}{\alpha} \fslow{\mathbf{u'}_\perp}}) \cdot \nabla_\perp + \textcolor{blue}{\frac{1}{\alpha} \fslow{w} \frac{\partial}{\partial z_s}} \right] \fslow{S \textcolor{blue}{- \tau T}} \\
    & \quad  + \fslow{w} = -\left( \nabla_\perp^2 + \textcolor{blue}{\frac{1}{\alpha^2} \frac{\partial^2}{\partial z_s^2}} \right) \fslow{b},
\end{split}
\end{equation}

and
\begin{equation}
\begin{split}
    & \left[ \frac{1}{\alpha} \frac{\partial}{\partial t_s} + \frac{\partial}{\partial t_f} + (\fslow{\mathbf{U}_\perp} + \textcolor{blue}{\frac{1}{\alpha} \fslow{\mathbf{u'}_\perp}}) \cdot \nabla_\perp + \fslow{w} \left( \textcolor{blue}{\frac{1}{\alpha} \frac{\partial}{\partial z_s}} + \frac{\partial}{\partial z_f} \right) \right] (\ffast{S} \textcolor{blue}{- \tau \ffast{T}}) \\
    & \quad + \left[ \ffast{\mathbf{u}}_\perp \cdot \nabla_\perp + \ffast{w} \left( \textcolor{blue}{\frac{1}{\alpha} \frac{\partial}{\partial z_s}} + \frac{\partial}{\partial z_f} \right) \right] (\ffast{S} \textcolor{blue}{- \tau \ffast{T}}) - \fslow{\left[ \ffast{\mathbf{u}}_\perp \cdot \nabla_\perp + \ffast{w} \left( \textcolor{blue}{\frac{1}{\alpha} \frac{\partial}{\partial z_s}} + \frac{\partial}{\partial z_f} \right) \right] (\ffast{S} \textcolor{blue}{- \tau \ffast{T}})} \\
    & \quad + \left[ \ffast{\mathbf{u}}_\perp \cdot \nabla_\perp + \textcolor{blue}{\ffast{w} \frac{1}{\alpha} \frac{\partial}{\partial z_s}} \right] \fslow{S \textcolor{blue}{- \tau T}} + \ffast{w} \\
    & \qquad = -\left( \nabla_\perp^2 + \textcolor{blue}{\frac{1}{\alpha^2} \frac{\partial^2}{\partial z_s^2} + \frac{2}{\alpha} \frac{\partial^2}{\partial z_s \partial z_f}} + \frac{\partial^2}{\partial z_f^2} \right) \ffast{b}.
\end{split}
\end{equation}

\subsection{Horizontal momentum}
The fast average of the horizontal momentum equation is
\begin{equation}
\begin{split}
    & \textcolor{blue}{\R \frac{1}{\Sc} \fslow{\left[ \ffast{\mathbf{u}}_\perp \cdot \nabla_\perp + \ffast{w} \left( \frac{1}{\alpha} \frac{\partial}{\partial z_s} + \frac{\partial}{\partial z_f} \right) \right] \ffast{\mathbf{u}}_\perp}} \\
    & \quad \textcolor{blue}{+ \R \frac{1}{\Sc} \left[ \frac{1}{\alpha} \frac{\partial}{\partial t_s} + (\fslow{\mathbf{U}_\perp} + \frac{1}{\alpha} \fslow{\mathbf{u'}_\perp}) \cdot \nabla_\perp + \frac{1}{\alpha} \fslow{w} \frac{\partial}{\partial z_s} \right] (\fslow{\mathbf{U}_\perp} + \frac{1}{\alpha} \fslow{\mathbf{u}'_\perp})} \\
    & \qquad = - \frac{1}{\alpha} \nabla_\perp \fslow{p} + \left( \nabla_\perp^2 + \frac{1}{\alpha^2} \frac{\partial^2}{\partial z_s^2} \right) (\fslow{\mathbf{U}_\perp} + \frac{1}{\alpha} \fslow{\mathbf{u}'_\perp}),
\end{split}
\end{equation}
and the horizontal mean of it yields
\begin{equation}
    \frac{\partial}{\partial t_s} \fslow{\mathbf{U}_\perp} + \frac{\partial}{\partial z_s}\fslow{ \overline{ \ffast{\mathbf{u}}_\perp \ffast{w} }} + \textcolor{blue}{\frac{1}{\alpha}\frac{\partial}{\partial z_s} \overline{\fslow{\mathbf{u}'_\perp} \fslow{w}}} = \frac{\Sc}{\R} \frac{1}{\alpha} \frac{\partial^2}{\partial z_s^2} \fslow{\overline{\mathbf{u}_\perp}},
\end{equation}
where the first term on the left-hand side is brought into balance by declaring that $\fslow{\mathbf{U}_\perp}$ varies on a slow timescale, as described in the main text.

The fast-averaged part subtracted from the full equation is
\begin{equation}
\begin{split}
    & \textcolor{blue}{\R \frac{\tau}{\Pr}\left[ \frac{1}{\alpha} \frac{\partial}{\partial t_s} + \frac{\partial}{\partial t_f} + (\fslow{\mathbf{U}_\perp} + \frac{1}{\alpha} \fslow{\mathbf{u'}_\perp}) \cdot \nabla_\perp + \fslow{w} \left( \frac{1}{\alpha} \frac{\partial}{\partial z_s} + \frac{\partial}{\partial z_f} \right) \right] \ffast{\mathbf{u}}_\perp} \\
    & \quad \textcolor{blue}{+ \R \frac{\tau}{\Pr} \left[ \ffast{\mathbf{u}}_\perp \cdot \nabla_\perp + \ffast{w} \left( \frac{1}{\alpha} \frac{\partial}{\partial z_s} + \frac{\partial}{\partial z_f} \right) \right] \ffast{\mathbf{u}}_\perp - \R \frac{\tau}{\Pr} \fslow{\left[ \ffast{\mathbf{u}}_\perp \cdot \nabla_\perp + \ffast{w} \left( \frac{1}{\alpha} \frac{\partial}{\partial z_s} + \frac{\partial}{\partial z_f} \right) \right] \ffast{\mathbf{u}}_\perp}} \\
    & \quad \textcolor{blue}{+ \R \frac{\tau}{\Pr} \left[ \ffast{\mathbf{u}}_\perp \cdot \nabla_\perp + \ffast{w} \frac{1}{\alpha} \frac{\partial}{\partial z_s} \right] (\fslow{\mathbf{U}_\perp} + \frac{1}{\alpha} \fslow{\mathbf{u'}_\perp})} \\
    & \qquad = - \nabla_\perp \ffast{p} + \left( \nabla_\perp^2 + \textcolor{blue}{\frac{1}{\alpha^2} \frac{\partial^2}{\partial z_s^2} + \frac{2}{\alpha} \frac{\partial^2}{\partial z_s \partial z_f}} + \frac{\partial^2}{\partial z_f^2} \right) \ffast{\mathbf{u}}_\perp.
\end{split}
\end{equation}


\subsection{Vertical momentum}
The fast average of the vertical momentum equation is
\begin{equation}
\begin{split}
    & \textcolor{blue}{\R \frac{\tau}{\Pr} \fslow{\left[ \ffast{\mathbf{u}}_\perp \cdot \nabla_\perp + \ffast{w} \left( \frac{1}{\alpha} \frac{\partial}{\partial z_s} + \frac{\partial}{\partial z_f} \right) \right] \ffast{w}}} \\
    & \quad \textcolor{blue}{+ \R \frac{\tau}{\Pr} \left[ \frac{1}{\alpha} \frac{\partial}{\partial t_s} + (\fslow{\mathbf{U}_\perp} + \frac{1}{\alpha} \fslow{\mathbf{u'}_\perp}) \cdot \nabla_\perp + \frac{1}{\alpha} \fslow{w} \frac{\partial}{\partial z_s} \right] \fslow{w}} \\
    & \qquad = \textcolor{blue}{- \frac{1}{\alpha^2} \frac{\partial}{\partial z_s} \fslow{p}} + \fslow{b} + \left( \nabla_\perp^2 + \textcolor{blue}{\frac{1}{\alpha^2} \frac{\partial^2}{\partial z_s^2}} \right) \fslow{w},
\end{split}
\end{equation}
and the $\ffast{w}$ part is
\begin{equation}
\begin{split}
    & \textcolor{blue}{\R \frac{\tau}{\Pr}\left[ \frac{1}{\alpha} \frac{\partial}{\partial t_s} + \frac{\partial}{\partial t_f} + (\fslow{\mathbf{U}_\perp} + \frac{1}{\alpha} \fslow{\mathbf{u'}_\perp}) \cdot \nabla_\perp + \fslow{w} \left( \frac{1}{\alpha} \frac{\partial}{\partial z_s} + \frac{\partial}{\partial z_f} \right) \right] \ffast{w}} \\
    & \quad \textcolor{blue}{+ \R \frac{\tau}{\Pr} \left[ \ffast{\mathbf{u}}_\perp \cdot \nabla_\perp + \ffast{w} \left( \frac{1}{\alpha} \frac{\partial}{\partial z_s} + \frac{\partial}{\partial z_f} \right) \right] \ffast{w} - \R \frac{\tau}{\Pr} \fslow{\left[ \ffast{\mathbf{u}}_\perp \cdot \nabla_\perp + \ffast{w} \left( \frac{1}{\alpha} \frac{\partial}{\partial z_s} + \frac{\partial}{\partial z_f} \right) \right] \ffast{w}}} \\
    & \quad \textcolor{blue}{+ \R \frac{\tau}{\Pr} \left[ \ffast{\mathbf{u}}_\perp \cdot \nabla_\perp + \ffast{w} \frac{1}{\alpha} \frac{\partial}{\partial z_s} \right] \fslow{w}} \\
    & \qquad = - \left( \textcolor{blue}{\frac{1}{\alpha} \frac{\partial}{\partial z_s}} + \frac{\partial}{\partial z_f} \right) \ffast{p} + \ffast{b} + \left( \nabla_\perp^2 + \textcolor{blue}{\frac{1}{\alpha^2} \frac{\partial^2}{\partial z_s^2} + \frac{2}{\alpha} \frac{\partial^2}{\partial z_s \partial z_f}} + \frac{\partial^2}{\partial z_f^2} \right) \ffast{w}.
\end{split}
\end{equation}

\subsection{Continuity}
Fast average of continuity equation
\begin{equation}
    0 = \nabla_\perp \cdot \fslow{\mathbf{u}'_\perp} + \frac{\partial}{\partial z_s} \fslow{w},
\end{equation}
and subtracting this off the total yields
\begin{equation}
    0 = \nabla_\perp \cdot \ffast{\mathbf{u}}'_\perp + \textcolor{blue}{\frac{1}{\alpha} \frac{\partial}{\partial z_s} \ffast{w}} + \frac{\partial}{\partial z_f} \ffast{w}.
\end{equation}

In the relevant limit ($\tau \ll 1$, $\Sc \gg 1$, $\R \ll 1$, $\alpha \gg 1$), the terms in blue above are subdominant, and the remaining terms form the closed system discussed in the main text, see Eqns.~(4.8)-(4.13).

\section{Simulation data}
Table 1 shows time-averaged quantities from simulations of the full model that are plotted in Fig.~4 in the main text. Note that we do not report anisotropy for the two smallest values of $\R$ because, at these small values of $\R$, the vertical scale of the mean flow is sufficiently large that we are unable to confirm that it is not constrained by our domain height $L_z$.

\begin{table}
\begin{center}
\begin{tabular}{c c c c c c c c c} 
\hline
$\R$ & $\mathbf{u}'_{\perp, \mathrm{rms}}$ & $\mathbf{U}_{\perp, \mathrm{rms}}$ & $w_\mathrm{rms}$ & anisotropy $\alpha$ & $S_\mathrm{rms}$ & $T_\mathrm{rms}$ & $|F_S|$ & $|F_T|$ \\ 
\hline
6.25e-03 & 1.48e-02 & 1.19e-05 & 4.41e-02 & --- & 7.18e-01 & 7.15e-01 & 2.93e-02 & 2.92e-02 \\
1.25e-02 & 3.20e-02 & 6.65e-05 & 1.03e-01 & --- & 1.23e+00 & 1.22e+00 & 1.15e-01 & 1.14e-01 \\
2.50e-02 & 6.97e-02 & 1.80e-03 & 2.72e-01 & 3.20e+01 & 2.30e+00 & 2.26e+00 & 5.62e-01 & 5.49e-01 \\ 
5.00e-02 & 1.45e-01 & 1.06e-02 & 7.32e-01 & 2.13e+01 & 4.42e+00 & 4.29e+00 & 2.92e+00 & 2.79e+00 \\ 
7.50e-02 & 2.02e-01 & 4.57e-02 & 1.24e+00 & 1.60e+01 & 6.22e+00 & 5.95e+00 & 6.94e+00 & 6.51e+00 \\ 
1.00e-01 & 2.59e-01 & 1.05e-01 & 1.77e+00 & 1.28e+01 & 7.88e+00 & 7.43e+00 & 1.26e+01 & 1.15e+01 \\ 
1.25e-01 & 3.09e-01 & 1.87e-01 & 2.29e+00 & 1.07e+01 & 9.32e+00 & 8.67e+00 & 1.92e+01 & 1.73e+01 \\ 
1.50e-01 & 3.70e-01 & 2.39e-01 & 2.77e+00 & 1.07e+01 & 1.06e+01 & 9.72e+00 & 2.66e+01 & 2.35e+01 \\ 
2.00e-01 & 4.88e-01 & 4.28e-01 & 3.57e+00 & 9.14e+00 & 1.23e+01 & 1.10e+01 & 4.00e+01 & 3.43e+01 \\ 
5.00e-01 & 1.35e+00 & 1.14e+00 & 8.01e+00 & 6.40e+00 & 2.19e+01 & 1.77e+01 & 1.64e+02 & 1.21e+02 \\ 
1.00e+00 & 3.07e+00 & 3.34e+00 & 1.45e+01 & 5.33e+00 & 3.65e+01 & 2.63e+01 & 5.01e+02 & 3.24e+02 \\ 
4.00e+00 & 1.64e+01 & 9.35e+00 & 4.69e+01 & 4.00e+00 & 1.20e+02 & 7.39e+01 & 5.12e+03 & 2.82e+03\\
\hline
\end{tabular}
\end{center}
\caption{Simulation results---data corresponding to simulations from the full equations shown in Fig.~4 in the main text. Here, ``anisotropy" refers to the blue crosses in the corresponding panel of Fig.~4 in the main text.}
\end{table}

\section{Convergence studies}

\begin{figure}
    \centering
    \includegraphics[width=0.9\textwidth]{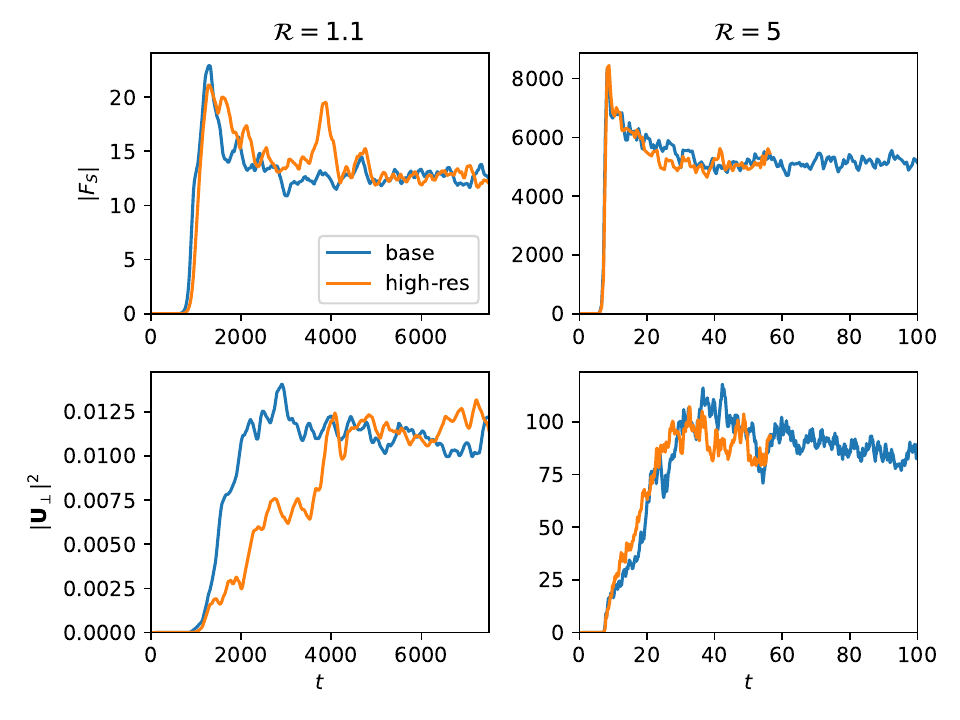}
    \caption{Salinity flux (top row) and kinetic energy in the mean flow (bottom row) for simulations with $\mathcal{R} = 1.1$ (left column; the same simulation as Fig.~1(d)-(f) in the manuscript) and $\mathcal{R} = 5$ (right column). In each panel, orange curves correspond to simulations run at twice the resolution in each direction compared to the blue curves: the $\mathcal{R} = 1.1$ simulation uses 64x32x512 grid points (with a domain size of 8x4x64 in units of $2\pi/\hat{k}_\mathrm{opt}$) for the base case and 128x64x1024 for the high-resolution case, while at $\mathcal{R} = 5$ (which uses a smaller $L_z$ as explained in the main text---domain size is 8x4x16 in units of $2\pi/\hat{k}_\mathrm{opt}$) we use 128x64x256 for the base case and 256x128x512 for the high-resolution case. For $\mathcal{R} = 5$, the high-resolution run was evolved for a shorter time to conserve computational resources. The mean flow is highly sensitive to the particular choice of initial conditions, and thus the high-resolution $\mathcal{R} = 1.1$ case does not immediately follow the same trend as the base case, though it eventually reaches the same mean flow amplitude in the saturated state.
    }
    \label{fig:ref1_conv_check}
\end{figure}

In Fig.~\ref{fig:ref1_conv_check}, we demonstrate for two of the cases reported in the main text ($\Ra = 1.1$ and $\Ra = 5$) that the salinity flux and the mean flow amplitude at saturation are minimally affected by a doubling of the resolution along each axis.

\bibliographystyle{jfm}
\bibliography{small-tau_salt-fingers}

\end{document}